\documentclass[prd,twocolumn,showpacs,preprintnumbers,amsmath,amssymb,superscriptaddress,floatfix,nofootinbib]{revtex4-2}

\usepackage{graphicx}
\usepackage{bm}
\usepackage{amsmath}
\usepackage{amsfonts}
\usepackage{amssymb}
\usepackage{color}
\usepackage{multirow}
\usepackage{subfigure}
\usepackage{ulem}
\usepackage[colorlinks, citecolor=blue,anchorcolor=red,menucolor=red, linkcolor=red,filecolor=red,runcolor=red,urlcolor=blue,frenchlinks=red]{hyperref}

\begin{document}
\title{Evidence of the low-lying baryon $\Sigma^*(1/2^-)$ in the process $\Lambda_c^+\to \eta\pi^+\Lambda$}

\author{Wen-Tao Lyu}
\affiliation{School of Physics, Zhengzhou
	University, Zhengzhou 450001, China}

\author{Sheng-Chao Zhang}
\affiliation{Joint Research Center for Theoretical Physics, School of Physics and Electronics, Henan University, Kaifeng 475004, China}\vspace{0.5cm}

\author{Guan-Ying Wang}
\email{wangguanying@henu.edu.cn}
\affiliation{Joint Research Center for Theoretical Physics, School of Physics and Electronics, Henan University, Kaifeng 475004, China}\vspace{0.5cm}

\author{Jia-Jun Wu}
\email{wujiajun@ucas.ac.cn}
\affiliation{School of Physical Sciences, University of Chinese Academy of Sciences, Beijing 100049, China}
\affiliation{Southern Center for Nuclear-Science Theory (SCNT), Institute of Modern Physics, Chinese Academy of Sciences, Huizhou 516000, China}\vspace{0.5cm}

\author{En Wang}\email{wangen@zzu.edu.cn (Corresponding author)}
\affiliation{School of Physics, Zhengzhou
	University, Zhengzhou 450001, China}
\affiliation{Guangxi Key Laboratory of Nuclear Physics and Nuclear Technology, Guangxi Normal University, Guilin 541004, China}\vspace{0.5cm}

\author{Li-Sheng Geng} \email{lisheng.geng@buaa.edu.cn}
\affiliation{School of Physics, Beihang University, Beijing 102206, China}
\affiliation{Peng Huanwu Collaborative Center for Research and Education, Beihang University, Beijing 100191, China}
\affiliation{Beijing Key Laboratory of Advanced Nuclear Materials and Physics, Beihang University, Beijing 102206, China }
\affiliation{Southern Center for Nuclear-Science Theory (SCNT), Institute of Modern Physics, Chinese Academy of Sciences, Huizhou 516000, China}
\vspace{0.5cm}

\author{Ju-Jun Xie} \email{xiejujun@impcas.ac.cn}
\affiliation{Southern Center for Nuclear-Science Theory (SCNT), Institute of Modern Physics, Chinese Academy of Sciences, Huizhou 516000, China}
\affiliation{Institute of Modern Physics, Chinese Academy of
Sciences, Lanzhou 730000, China} \affiliation{School of Nuclear Sciences and Technology, University of Chinese Academy of Sciences, Beijing 101408, China} 
\vspace{0.5cm}

\begin{abstract}
Motivated by the Belle measurements of the process $\Lambda_c^+\to \eta\pi^+\Lambda$, we investigate this process by considering the contributions from the $\Lambda(1670)$, $a_0(980)$, and $\Sigma(1385)$. In addition, we also consider the predicted low-lying baryon $\Sigma^*(1/2^-)$. 
Our results involving the $\Sigma^*(1/2^-)$ are favored by fitting to the Belle data of the $\eta\Lambda$ and $\pi^+\Lambda$ invariant mass distributions. Furthermore, we predict the $\eta\pi^+$ invariant mass distribution and the angular distribution $d\Gamma/d{\rm cos}\theta$, which are significantly different depending on whether or not the contribution from the $\Sigma^*(1/2^-)$ is considered.  Finally, we show that, with the contribution from the $\Sigma^*(1/2^-)$, the calculated Dalitz plot agrees with the Belle measurements.
Future precise measurements of the process $\Lambda_c^+\to \eta\pi^+\Lambda$ could shed further light on the existence of the low-lying $\Sigma^*(1/2^-)$. 
\end{abstract}

\date{\today}

\maketitle

\section{Introduction}\label{sec1}

Since the charmoniumlike state $X(3872)$ was observed in the $\pi^+\pi^-J/\psi$ invariant mass distribution of the process $B^{\pm}\to K^{\pm}\pi^+\pi^-J/\psi$ by the Belle Collaboration in 2003~\cite{Belle:2003nnu},  a plethora of hadrons beyond the conventional quark model have been observed experimentally. Exploring their nature is crucial to understanding the nonperturbative properties of quantum chromodynamics (QCD)~\cite{Chen:2016spr,Chen:2022asf,Guo:2017jvc,Oset:2016lyh,Liu:2024uxn,Gao:1999ar}. 
It should be stressed that most of these exotic hadrons are located near the mass thresholds of pairs of hadrons, and many theoretical studies have concluded that the hadron-hadron interactions play an essential role in these exotic hadrons~\cite{Guo:2017jvc,Oset:2016lyh,Liu:2024uxn}.

As we know, the low-lying baryon $\Lambda(1405)$ with $J^P=1/2^-$ could be generated from the $\bar{K}N$ interaction in the unitary chiral approach~\cite{Oset:1997it,Jido:2003cb}.
%
However, the low-lying excited $\Sigma^*(1/2^-)$ has not yet been well-established experimentally and theoretically~\cite{ParticleDataGroup:2022pth}.
Although the state $\Sigma(1620)$ with $J^P=1/2^-$ is listed in the Review of Particle Physics (RPP)~\cite{ParticleDataGroup:2022pth}, its status is only one star (meaning that the evidence of its existence is poor) and is omitted from the summary tables, which implies that this state still needs to be confirmed. 
Furthermore, the $\Sigma^*(1/2^-)$ with a mass around 1.4 GeV as a partner of $\Lambda(1405)$ is not found yet. 
Thus, searching for the state $\Sigma^*(1/2^-)$ is crucial to deepen our understanding of low-lying excited baryons~\cite{Crede:2013kia,Klempt:2009pi,Oset:2016lyh}, and many theoretical works have been devoted to this issue~\cite{Wang:2024jyk}.

For instance, the $S$-wave meson-baryon interaction in the $S=-1$ sector was studied in the unitary chiral approach, and one $\Sigma^*(1/2^-)$ with a mass around the $\Bar{K}N$ threshold was predicted in Refs.~\cite{Oset:2001cn,Oller:2000fj,Oset:1997it,Khemchandani:2018amu,Kamiya:2016jqc,Jido:2003cb,Oller:2006jw,Garcia-Recio:2002yxy,Lutz:2001yb}, which is supported by the analysis of the CLAS data on the process $\gamma p \to K\Sigma\pi$~\cite{CLAS:2013rjt,Roca:2013cca}. 
By analyzing the relevant data of $K^-p\to \Lambda\pi^+\pi^-$, Refs.~\cite{Wu:2009nw,Wu:2009tu} suggested that there may exist one $\Sigma^*(1/2^-)$ resonance with a mass about 1380~MeV. 
In the effective Lagrangian approach, Ref.~\cite{Gao:2010hy} found that the $\Sigma^*(1/2^-)$ plays an important role in the $K\Sigma^*(1385)$ photoproduction.  
In addition, it is also suggested to search for the $\Sigma^*(1/2-)$ in the processes of $\Lambda_c \to \eta\Lambda\pi^+$~\cite{Xie:2017xwx},  $\Lambda_c^+\to \pi^+\pi^-\pi^0\Sigma^+$~\cite{Xie:2018gbi},  $\chi_{c0}(1P) \to \bar{\Lambda}\Sigma\pi$~\cite{Liu:2017hdx}, $\chi_{c0}(1P) \to \bar{\Sigma}\Sigma\pi$~\cite{Wang:2015qta}, and $\gamma N\to K\Sigma^{*}$~\cite{Lyu:2023oqn,Kim:2021wov}. 
Recently, two $\Sigma^*(1/2^-)$ states were predicted in the global study of meson-baryon scattering in the covariant baryon chiral perturbation theory, including the next-to-next-to-leading order contributions, where the narrower pole located at $(1432,-i18)$~MeV shows up as a cusp, and the broader one located at $(1364,-i110)$~MeV becomes a broad enhancement in the real axis of the amplitude squared~\cite{Lu:2022hwm}.

Based on a 980~fb$^{-1}$ data sample, the Belle Collaboration measured the process $\Lambda^+_c\to \eta\pi^+\Lambda$, which had been measured by CLEO~\cite{CLEO:1995cbq} and BESIII~\cite{BESIII:2018qyg} earlier, and presented the $\eta\Lambda$ and $\pi^+\Lambda$ invariant mass distributions, which show significant signals of the resonances $\Lambda(1670)$, $a_0(980)$, and $\Sigma(1385)$~\cite{Belle:2020xku}. 
In Ref.~\cite{Wang:2022nac}, we have analyzed this process by considering the intermediate state $\Sigma(1385)$ ($J^P=3/2^+$), and the $S$-wave $\eta\Lambda$ and $\eta\pi^+$ final-state interactions in the chiral unitary approach, which dynamically generate the $\Lambda(1670)$ and $a_0(980)$ resonances. Our results could reproduce the Belle data of the $\pi^+\Lambda$ and $\eta\Lambda$ invariant mass distributions. 
However, the Dalitz plot of the process $\Lambda^+_c\to \eta\pi^+\Lambda$ has more yields appearing in the region of $M^2_{\Lambda\pi^+}<1.85$~GeV$^2$ and $M^2_{\eta\Lambda}<4$~GeV$^2$, which cannot be well-described in Ref.~\cite{Wang:2022nac}. 
The discrepancies may imply some contributions from the resonance $\Sigma^*(1/2^-)$~\cite{Xie:2017xwx}.
In principle, the signal of $\Sigma^*(1/2^-)$ could overlap with the one of $\Sigma(1385)$, and will be difficult to be directly observed in the $\Lambda\pi^+$ invariant spectrum. However, the contributions from $\Sigma^*$ is crucial to understanding the invariant mass spectrum of $\pi^+ \eta$ correctly.
Thus, the invariant mass distribution of $\pi^+ \eta$ is essential for us to determine the existence of $\Sigma^*(1/2^-)$, but it is missing in the publication of the Belle Collaboration~\cite{Belle:2020xku}.


In this work, besides the intermediate $\Sigma(1385)$, the $S$-wave final-state interactions of the $\eta\Lambda$ and $\eta\pi^+$, we also consider the contribution from the intermediate $\Sigma^*(1/2^-)$ in the process of $\Lambda^+_c\to \eta\pi^+\Lambda$ and show the evidence of this state by analyzing the Belle measurements. 

This paper is organized as follows. In Sec.~\ref{sec2}, we present the theoretical formalism for the process $\Lambda^+_c\to \eta\pi^+\Lambda$. Numerical results are presented in Sec.~\ref{sec3}. Finally, we present a summary in the last section.

\section{Formalism}\label{sec2}
 
This section will introduce the theoretical formalism for the process  $\Lambda^+_c\to \eta\pi^+\Lambda$. 
First, in Sec.~\ref{sec2a}, we present the mechanisms for this process via the $S$-wave meson-baryon (MB) interactions, which would generate the state $\Lambda(1670)$. 
Second, in Sec.~\ref{sec2b}, we present the mechanisms for this process via the meson-meson (MM) interactions, which would generate the scalar $a_0(980)$. 
The contributions from the intermediate states $\Sigma(1385)$ and $\Sigma^*(1/2^-)$ are discussed in Sec.~\ref{sec2c}. 
Finally, we present the formalism for the invariant mass distributions and angular distributions of the process $\Lambda^+_c\to \eta\pi^+\Lambda$ in Sec.~\ref{sec2e}.

\subsection{$S$-wave meson-baryon ($MB$) interactions and $\Lambda(1670)$} \label{sec2a}

\begin{figure}[htbp]
	\centering
	
	\includegraphics[scale=0.65]{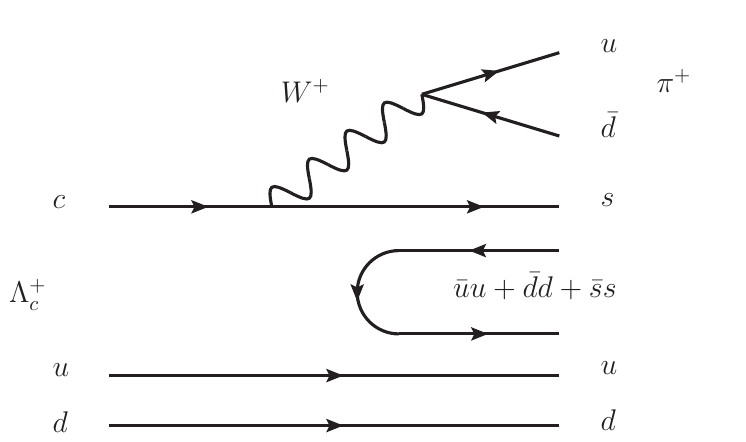}
	
	\caption{Quark-level diagram for the $\Lambda^+_c\to \pi^+s(\bar{u}u+\bar{d}d+\bar{s}s)ud$ process.}\label{fig:MB-quark}
\end{figure}

 For the process $\Lambda_c^+\to \pi^+ (MB) \to \pi^+ \eta \Lambda$, the dominant mechanism is the $W^+$ external emission, as shown in Fig.~\ref{fig:MB-quark}. 
 The $c$ quark of the initial $\Lambda^+_c$ will firstly weakly decay into an $s$ quark and a $W^+$ boson, then the $W^+$ boson subsequently decays into a $u\bar{d}$ quark pair. 
 The $u\bar{d}$ quark from the $W^+$ boson will hadronize into a $\pi^+$ meson of high momentum, while the $s$ quark from the initial $\Lambda^+_c$ and the $ud$ quark pair of the initial $\Lambda^+_c$, together with the quark pair $\bar{q}q=\bar{u}u+\bar{d}d+\bar{s}s$ created from the vacuum with the quantum numbers $J^{PC}=0^{++}$, hadronize into hadron pairs, which could be expressed as
\begin{eqnarray}
    \Lambda_c^+&\Rightarrow&\frac{1}{\sqrt{2}}c(ud-du) \Rightarrow \frac{1}{\sqrt{2}}W^+s(ud-du) \nonumber \\
    &\Rightarrow &\frac{1}{\sqrt{2}}u\bar{d}s(\bar{u}u+\bar{d}d+\bar{s}s)(ud-du) \nonumber \\
        &\Rightarrow&\frac{1}{\sqrt{2}}\pi^+s(\bar{u}u+\bar{d}d+\bar{s}s)(ud-du) \nonumber \\
 &\Rightarrow&\frac{1}{\sqrt{2}}\pi^+\sum M_{3i} q_i(ud-du).
\end{eqnarray}
Here  $M$ is the matrix of the pseudoscalar mesons,
\begin{eqnarray}
		M =\left(\begin{matrix}   \frac{{\eta}}{\sqrt{3}}+ \frac{{\pi}^0}{\sqrt{2}} +\frac{{\eta}'}{\sqrt{6}}  & \pi^+  & K^{+}  \\
			\pi^-  &    \frac{{\eta}}{\sqrt{3}}- \frac{{\pi}^0}{\sqrt{2}} +\frac{{\eta}'}{\sqrt{6}}  &  K^{0} \\
			K^{-}  &  \bar{K}^{0}   & \sqrt{\frac{2}{3}}{\eta}'  -\frac{{\eta}}{\sqrt{3}}
		\end{matrix}
		\right),
	\end{eqnarray}
 where we have considered the approximate $\eta-\eta'$ mixing~\cite{Bramon:1992kr,Lyu:2023ppb}.
 
Then, we have the components of the pseudoscalar meson and octet baryon as
\begin{widetext}
\begin{eqnarray}
 \frac{1}{\sqrt{2}} \sum M_{3i} q_i(ud-du) &=& K^- \left[ \frac{1}{\sqrt{2}}u(ud-du) \right]  + \bar{K}^0 \left[ \frac{1}{\sqrt{2}}d(ud-du) \right] -\frac{{\eta}}{\sqrt{3}}  \left[ \frac{1}{\sqrt{2}}s(ud-du)  \right] \nonumber \\
&=&K^- p+\bar{K}^0 n +\frac{\sqrt{2}}{3}\eta \Lambda, \label{eq:MBchannel}
\end{eqnarray}
\end{widetext}
where we have ignored the $\eta' \Lambda$ channel since its threshold ($2073.5$~MeV) is far from the mass region of $\Lambda(1670)$. 
Here, we take the flavor wave functions for the octet baryons as~\cite{Pavao:2017cpt,Miyahara:2016yyh,Lyu:2023aqn}
\begin{gather}
		p=	\frac{u(ud-du)}{\sqrt{2}}, ~~~		n=\frac{d(ud-du)}{\sqrt{2}}, \\
		\Lambda=\frac{u(ds-sd)+d(su-us)-2s(ud-du)}{2\sqrt{3}}.
\end{gather}

\begin{figure}[htbp]
		\centering
	\subfigure[]{
		\includegraphics[scale=0.65]{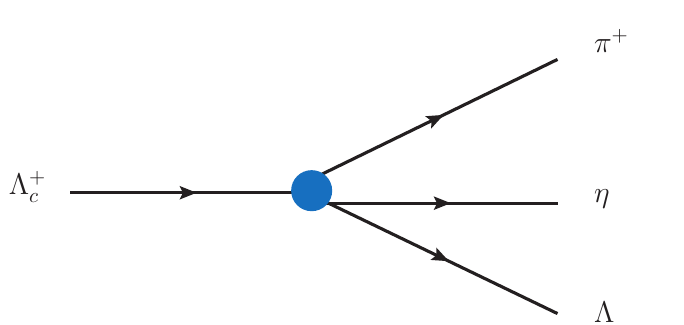}
	}
	
	\subfigure[]{
		\includegraphics[scale=0.65]{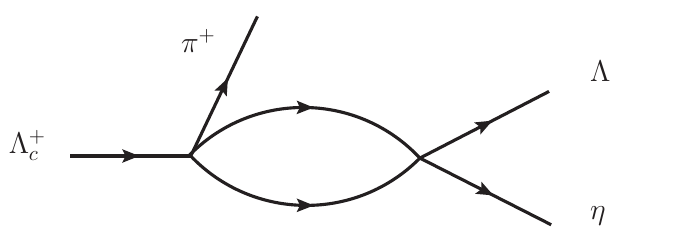}
	}
	\caption{The mechanisms for the $\Lambda_c\to \pi^+\eta\Lambda$ decay. (a) tree diagram, (b) the $S$-wave meson-baryon interactions. 
		}\label{fig:MB-hadron}
\end{figure}

Then, the process $\Lambda^+_c\to\eta\pi^+\Lambda$ decay could occur via the tree diagram of Fig.~\ref{fig:MB-hadron}(a), and the $S$-wave meson-baryon interaction of Fig.~\ref{fig:MB-hadron}(b). The decay amplitude could be expressed as	
\begin{equation}\label{t-MB}
	\begin{aligned}
		\mathcal{T}^{MB}(M_{\eta\Lambda}) = & V_p \left[ \frac{\sqrt{2}}{3}+G_{K^-p}(M_{\eta\Lambda})t_{K^-p\to\eta\Lambda}(M_{\eta\Lambda}) \right.\\
		&+G_{\bar{K}^0n}(M_{\eta\Lambda})t_{\bar{K}^0n\to\eta\Lambda}(M_{\eta\Lambda})  \\
		&\left. +\frac{\sqrt{2}}{3}G_{\eta\Lambda}(M_{\eta\Lambda})t_{\eta\Lambda\to\eta\Lambda}(M_{\eta\Lambda}) \right], 
	\end{aligned}
\end{equation}		
where $G_{l}$ are the loop functions of the meson-baryon system, and the parameter $V_p$, accounting for the weak decay and hadronization strength, is approximately independent of the final-state interactions~\cite{Roca:2015tea,Oset:2016lyh}. 
The $t_{MB\to\eta\Lambda}$ are the transition amplitudes, which can be obtained by solving the Bethe-Salpeter equation,
\begin{equation}\label{BS}
	T=[1-VG]^{-1}V.
\end{equation}
The transition potential $V_{ij}$ is taken from Ref.~\cite{Oset:2001cn},
\begin{equation}\label{V}
	\begin{aligned}
		V_{ij} &=-C_{ij}\frac{1}{4f^2}(2\sqrt{s}-M_i-M_j)\\
		&\times\left(\dfrac{M_i+E_i}{2M_i}\right)^{1/2}\left(\dfrac{M_j+E_j}{2M_j}\right)^{1/2},
	\end{aligned}
\end{equation}
where $E_i$ and $M_i$ are the energy and mass of the baryon in the $i$th channel, and the coefficients $C_{ij}$ reflect the $SU(3)$ flavor symmetry, and are given in Table~I of Ref.~\cite{Oset:1997it}. 
$\sqrt{s}$ is the invariant mass of the $\eta\Lambda$ system. 
The coupling $f$ is the pseudoscalar decay constant, for which we use,
\begin{equation}
	f=1.15f_{\pi},~~f_{\pi}=93~\mathrm{MeV}.
\end{equation}
The $G_{l}$ can be written by~\cite{Oset:2001cn}
\begin{equation}\label{G-DR}
	\begin{aligned}
		G_{l}  =&i \int \frac{d^4 q}{(2 \pi)^4}\frac{M_l}{E_l(\vec{q}\,)} \frac{1}{\sqrt{s}-q^0-E_l(\vec{q}\,)+i \epsilon} \frac{1}{q^2-m_l^2+i \epsilon} \\
		 =&\frac{2M_l}{16 \pi^2}\left\{a_l(\mu)+\ln \frac{M_l^2}{\mu^2}+\frac{s+m_l^2-M_l^2}{2 s} \ln \frac{m_l^2}{M_l^2}\right. \\
		& +\frac{|\vec{q}\,|}{\sqrt{s}}\left[\ln \left(s-\left(M_l^2-m_l^2\right)+2 |\vec{q}\,| \sqrt{s}\right)\right. \\
		& +\ln \left(s+\left(M_l^2-m_l^2\right)+2 |\vec{q}\,| \sqrt{s}\right) \\
		& -\ln \left(-s+\left(M_l^2-m_l^2\right)+2 |\vec{q}\,| \sqrt{s}\right) \\
		& \left.\left.-\ln \left(-s-\left(M_l^2-m_l^2\right)+2 |\vec{q}\,| \sqrt{s}\right)\right]\right\},
	\end{aligned}
\end{equation}
where $m_l$ and $M_l$ are the masses of meson and baryon of the $l$th channel. 
Here, we take $\mu=630$~MeV. 
Since the pole position of $\Lambda(1670)$ is quite sensitive to the value of $a_{K\Xi}$ and only moderately sensitive to $a_{\bar{K}N}$, $a_{\pi\Sigma}$, $a_{\eta\Lambda}$, we take $a_{K\Xi}$ to be a free parameter, and adopt the values $a_{\bar{K}N}=-1.84$, $a_{\pi\Sigma}=-2.00$, and $a_{\eta\Lambda}=-2.25$ from Ref.~\cite{Oset:2001cn}.

\subsection{$S$-wave meson-meson ($MM$) interactions and $a_0(980)$} \label{sec2b}

\begin{figure}[htbp]
		\centering
	
	\includegraphics[scale=0.65]{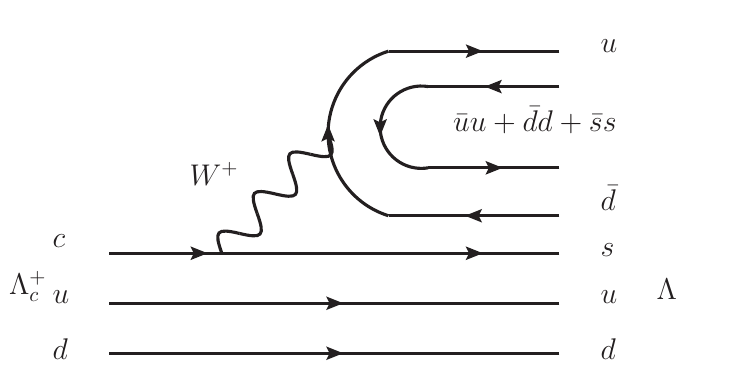}
	
	\caption{ Quark level diagram for the $\Lambda^+_c\to \Lambda u(\bar{u}u+\bar{d}d+\bar{s}s)\bar{d}$ process.}\label{fig:MM-quark}
\end{figure}

The primary mechanism for the process $\Lambda_c^+\to MM \Lambda \to \pi^+\eta \Lambda$ is the $W^+$ external emission, as shown in Fig.~\ref{fig:MM-quark}. 
The $c$ quark of the initial $\Lambda^+_c$ weakly decays into an $s$ quark and a $W^+$ boson, and then the $W^+$ boson subsequently decays into a $u\bar{d}$ quark pair. 
The $s$ quark from the $\Lambda^+_c$ decay and the $ud$ quark pair of the initial $\Lambda^+_c$ will hadronize into a $\Lambda$ baryon, while the $u\bar{d}$ quark pair from the $W^+$ boson, together with the quark pair $\bar{q}q=\bar{u}u+\bar{d}d+\bar{s}s$ created from the vacuum with the quantum numbers $J^{PC}=0^{++}$, hadronize into hadron pairs, 
\begin{eqnarray}
    \Lambda_c^+&\Rightarrow&\frac{1}{\sqrt{2}}c(ud-du) \nonumber \\
    &\Rightarrow&\frac{1}{\sqrt{2}}W^+s(ud-du) \nonumber \\
    &\Rightarrow &\frac{1}{\sqrt{2}}u\bar{d}s(ud-du) \nonumber \\
        &\Rightarrow&\frac{1}{\sqrt{2}}u(\bar{u}u+\bar{d}d+\bar{s}s)\bar{d}s(ud-du)   \nonumber \\
 &\Rightarrow&\frac{1}{\sqrt{2}}\sum M_{1i} M_{i2} s(ud-du) \nonumber \\
 &\Rightarrow& \left( \frac{2}{\sqrt{3}}\pi^+\eta+K^+\bar{K}^0  \right)  \frac{\sqrt{6}}{3} \Lambda \nonumber \\
 &=& \left( \frac{2\sqrt{2}}{3}\pi^+\eta+\frac{\sqrt{6}}{3}K^+\bar{K}^0  \right)   \Lambda.
\end{eqnarray}

\begin{figure}[htbp]
	\subfigure[]{
	\includegraphics[scale=0.65]{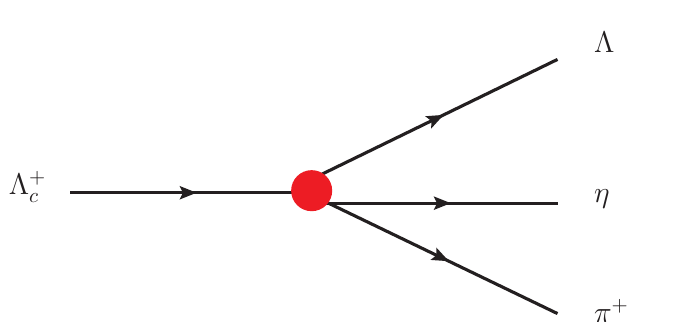}
}

\subfigure[]{
	\includegraphics[scale=0.65]{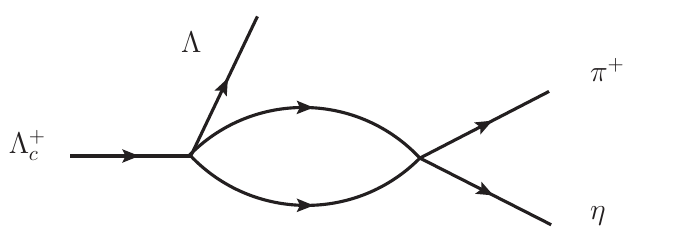}
}
	\caption{The mechanisms for the $\Lambda^+_c\to \pi^+\eta\Lambda$ decay. (a) tree diagram, (b) the $S$-wave pseudoscalar meson-meson interactions}\label{fig:MM-hadron}
\end{figure}

Then, the process $\Lambda^+_c\to\eta\pi^+\Lambda$ decay occurs via the tree diagram of Fig.~\ref{fig:MM-hadron}(a), and the $S$-wave pseudoscalar meson-pseudoscalar meson interaction of Fig.~\ref{fig:MM-hadron}(b). The amplitude could be expressed as
\begin{equation}\label{t-MM}
	\begin{aligned}
		\mathcal{T}^{MM} =&V'_p \left[\frac{2\sqrt{2}}{3}+\frac{2\sqrt{2}}{3}\tilde{G}_{\pi^+\eta}t_{\pi^+\eta\to\pi^+\eta} \right. \\
		&\left. +\frac{\sqrt{6}}{3}\tilde{G}_{K^+\bar{K}^0}t_{K^+\bar{K}^0\to\pi^+\eta}\right],
	\end{aligned}
\end{equation}
where $\tilde{G}_{MM}$ in Eq.~(\ref{t-MM}) is the loop function of the meson-meson system, and $t_{MM\to\pi^+\eta}$ are the scattering matrices of the coupled channels. 
The transition amplitude of $t_{MM\to\pi^+\eta}$ is obtained in the chiral unitary approach. The details can be found in Ref.~\cite{Wang:2022nac}. 
Here, $V'_p$ accounts for the strength of the weak vertex and hadronization. Note that the tree diagrams of Figs.~\ref{fig:MB-hadron}(a) and \ref{fig:MM-hadron}(a) are similar, but their vertices are different.
In this work, we take $V'_p=0.155 V_p$, which is obtained assuming that the amplitudes of Eq.~(\ref {t-MB}) and  Eq.~(\ref {t-MM}) give rise to the same width for the process $\Lambda_c\to \eta\pi^+\Lambda$~\cite{Xie:2016evi}.
In this work, the loop function $\tilde{G}_l$  in Eq.~(\ref{t-MM}) is calculated using the cutoff method,
\begin{equation}\label{Gb}
			\begin{aligned}
				\tilde{G}_{l} & =i \int \frac{d^4 q}{(2 \pi)^4} \frac{1}{(p-q)^2-m_2^2+i \epsilon} \frac{1}{q^2-m_1^2+i \epsilon} \\
				& =\int_0^{q_{\rm max}}  \frac{d^3 q}{(2 \pi)^3}  \frac{\omega_1+\omega_2}{\omega_1\omega_2}\frac{1}{s-(\omega_1+\omega_2)^2+i \epsilon},
			\end{aligned}
		\end{equation}
with $\omega_{1/2}=\sqrt{m^2_{1/2}+\vec{q}^{\,2}}$. 
In this work, we take the cutoff momentum $q_{\rm max}=600$~MeV~\cite{Xie:2014tma,Wang:2021naf,Feng:2020jvp}.
It should be stressed that, in this work, we use the $G$ function of the dimensional regularization method, as shown in Eq.~(\ref{G-DR}), for the $S$-wave meson-baryon system, and the one of the cut-off method, as shown in Eq.~(\ref{Gb}), for the $S$-wave meson-meson system, consistent with the previous works~\cite{Oset:2001cn,Xie:2016evi,Wang:2022nac,Xie:2014tma}.

\subsection{Intermediate states $\Sigma(1385)$ and $\Sigma^*(1/2^-)$} \label{sec2c}

\begin{figure}[htbp]
	\centering
	\includegraphics[scale=0.65]{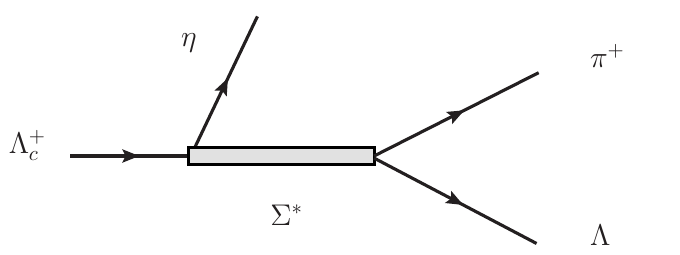}
	\caption{The contribution from the intermediate $\Sigma^*$.}\label{fig:Sigmastar}
\end{figure}

\begin{figure}[htbp]
	\centering
	\includegraphics[scale=0.40]{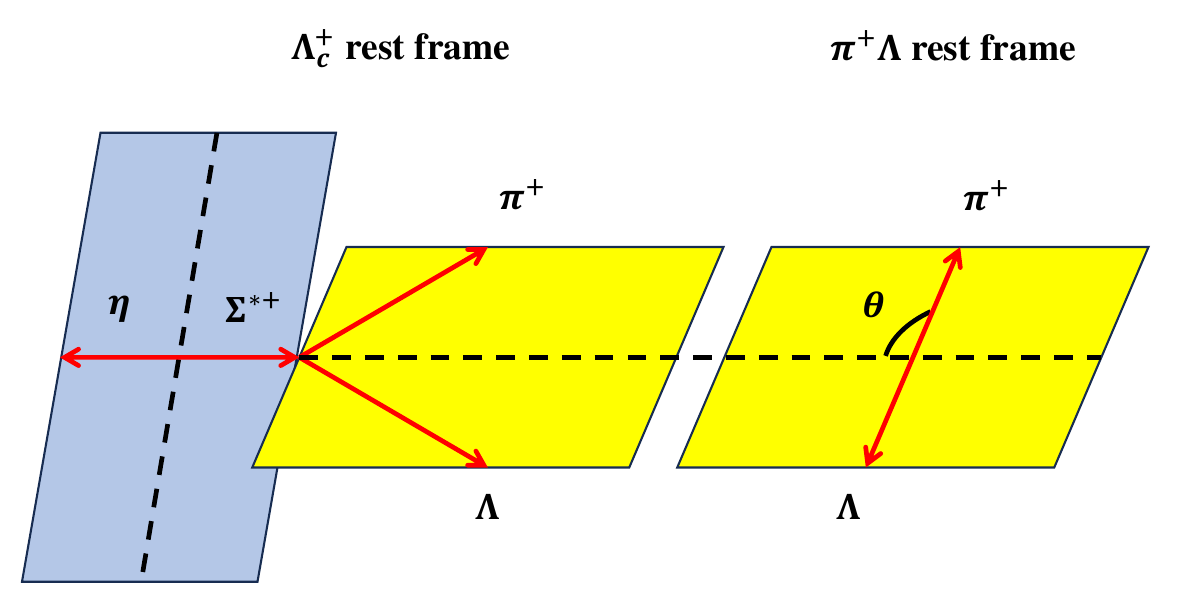}
	\caption{The angle $\theta$ between $\pi^+$ and $\eta$ in the center-of-mass frame of the $\pi^+\Lambda$ rest frame.}\label{fig:angle-diagram}
\end{figure}

For the process $\Lambda^+_c\to \eta\Sigma(1385)\to \eta\pi^+\Lambda$  depicted in Fig.~\ref{fig:Sigmastar}, the  amplitude can be written as follows~\cite{Wang:2015pcn},
\begin{equation}\label{t1385}
	\mathcal{T}^{\Sigma(1385)} =\dfrac{V^{\Sigma(1385)}|p^*_{\pi^+}||p^*_{\eta}|{\rm cos}\theta}{M^2_{\pi^+\Lambda}-M^2_{\Sigma(1385)}+iM_{\Sigma(1385)}\Gamma_{\Sigma(1385)}}  ,
\end{equation}
where $V^{\Sigma(1385)}$ is the relative strength of the contribution from the intermediate resonance $\Sigma(1385)$, and $|p^*_{\pi}|$ and $|p^*_{\eta}|$ are the momenta of $\pi^+$ and $\eta$ in the rest frame of the $\pi^+\Lambda$ system, respectively, and $\theta$ is the angle between $\pi^+$ and $\eta$ in the center-of-mass frame of the $\pi^+\Lambda$ system, as depicted in Fig.~\ref{fig:angle-diagram}, which are given by~\cite{Wang:2015pcn}
\begin{equation}\label{ppi}
	|p^*_{\pi^+}|=\dfrac{\lambda^{1/2}(M_{\pi^+\Lambda}^2,m_{\pi^+}^2,M_{\Lambda}^2)}{2M_{\pi^+\Lambda}},
\end{equation}
\begin{equation}\label{pds}
	|p^*_{\eta}|=\dfrac{\lambda^{1/2}(M_{\Lambda_c}^2,m_{\eta}^2,M_{\pi^+\Lambda}^2)}{2M_{\pi^+\Lambda}},
\end{equation}
\begin{equation}\label{theta}
	{\rm cos}\theta=\dfrac{M_{\eta\Lambda}^2-M_{\Lambda^+_c}^2-m_{\pi^+}^2+2P_{\Lambda^+_c}^0P_{\pi^+}^0}{2|\vec{p}_{\pi^+}||\vec{p}_{\eta}|},
\end{equation}
with the K$\ddot{\text{a}}$llen function $\lambda(x,y,z)=x^2+y^2+z^2-2xy-2yz-2xz$.
In the $\pi^+\Lambda$ rest frame, $\vec{p}_{\Lambda^+_c}=\vec{p}_{\eta}$, $\vec{p}_{\pi^+}=-\vec{p}_{\Lambda}$, and the $\Lambda_c^+$ and $\pi^+$ energies are
\begin{equation}
	\begin{aligned}
		P_{\Lambda^+_c}^0&=\sqrt{M_{\Lambda^+_c}^2+|\vec{p}_{\Lambda^+_c}|^2}=\sqrt{M_{\Lambda^+_c}^2+|\vec{p}_{\eta}|^2}, \nonumber\\
		P_{\pi^+}^0&=\sqrt{m_{\pi^+}^2+|\vec{p}_{\pi^+}|^2}.
	\end{aligned}
\end{equation}


As we discussed in the introduction, besides the $\Sigma(1385)$, we will also take into account the contribution from the resonance $\Sigma^*(1/2^-)$ in the process $\Lambda^+_c\to \eta\Sigma^*(1/2^-)\to \eta\pi^+\Lambda$, as shown in Fig.~\ref{fig:Sigmastar}.
Since the state $\Sigma^*(1/2^-)$ has not been confirmed by experiments, in this work we take a simple Breit-Wigner form for the $\Sigma^*(1/2^-)$ contribution, rather than the more complex form, such as the energy dependence of the coupling and widths, in order to reduce the number of free parameters. One should be noted that the line shape of the $\Sigma^*(1/2^-)$ depends on the adopted amplitude form, and could affect the fitted results.
The  amplitude of Fig.~\ref{fig:Sigmastar} can be written as follows,
\begin{equation}\label{t1380}
	\mathcal{T}^{\Sigma^*(1/2^-)} =\dfrac{V^{\Sigma^*(1/2^-)}M_{\Sigma^*(1/2^-)}\Gamma_{\Sigma^*(1/2^-)}}{M^2_{\pi^+\Lambda}-M^2_{\Sigma^*(1/2^-)}+iM_{\Sigma^*(1/2^-)} \Gamma_{\Sigma^*(1/2^-)}}  ,
\end{equation}
where $V^{\Sigma^*(1/2^-)}$ is the relative strength of the contribution from the intermediate resonance $\Sigma^*(1/2^-)$.
The predicted mass of the $\Sigma^*(1/2^-)$ stat is about 1380~MeV~ and 1430~MeV.
 Since there is no structure around 1430~MeV in the $\pi^+\Lambda$ invariant mass distribution of the Belle data,  if the $\Sigma^*(1/2^-)$ exists, one expects that its mass should be close to the one of $\Sigma(1385)$. To reduce the number of free parameters, we take its predicted mass $M_{\Sigma^*(1/2^-)}=1380$~MeV and width  $\Gamma_{\Sigma^*(1/2^-)}=120$~MeV as in Refs.~\cite{Wu:2009nw,Wu:2009tu,Liu:2017hdx}.

\subsection{Invariant mass distributions} \label{sec2e}
With all the ingredients obtained in the previous section, one can write down the modulus squared of the total amplitude as,
\begin{equation}\label{ttotal}
	\begin{aligned}
	|\mathcal{T}^{\text {Total}}|^2=&|\mathcal{T}^{MM}+ \mathcal{T}^{MB}e^{i\phi} +\mathcal{T}^{\Sigma(1385)}e^{i\phi^{\prime}} \\
	&+\mathcal{T}^{\Sigma^*(1/2^-)}e^{i\phi^{\prime\prime}}|^2,
	\end{aligned}
\end{equation}
where the $\phi$, $\phi^{\prime}$, and $\phi^{\prime\prime}$ are the phase angles between different contributions.
Then, the invariant mass distribution of the double differential width of the $\Lambda^+_c\to \eta\pi^+\Lambda$ can be written as
\begin{eqnarray}\label{dwidth}
	\dfrac{d^2\Gamma}{dM_{\pi^+\eta}^2dM_{\eta\Lambda}^2} &=& \frac{1}{(2\pi)^3}\dfrac{M_{\Lambda}}{8M_{\Lambda^+_c}^2}|\mathcal{T}^{\text {Total}}|^2, \\
 	\dfrac{d^2\Gamma}{dM_{\pi^+\eta}^2dM_{\pi^+\Lambda}^2} &=& \frac{1}{(2\pi)^3}\dfrac{M_{\Lambda}}{8M_{\Lambda^+_c}^2}|\mathcal{T}^{\text {Total}}|^2.
\end{eqnarray}
For a given value of $M_{12}$, the range of $M_{23}$ is determined by~\cite{ParticleDataGroup:2022pth},
\begin{align}
	&\left(m_{23}^2\right)_{\min}=\left(E_2^*+E_3^*\right)^2-\left(\sqrt{E_2^{* 2}-m_2^2}+\sqrt{E_3^{* 2}-m_3^2}\right)^2, \nonumber\\
	&\left(m_{23}^2\right)_{\max}=\left(E_2^*+E_3^*\right)^2-\left(\sqrt{E_2^{* 2}-m_2^2}-\sqrt{E_3^{* 2}-m_3^2}\right)^2, 
\end{align}
where $E_2^{*}$ and $E_3^{*}$ are the energies of particles 2 and 3 in the $M_{12}$ rest frame, which are written as
\begin{align}
	&E_2^{*}=\dfrac{M_{12}^2-m_1^2+m_2^2}{2M_{12}}, \nonumber\\
	&E_3^{*}=\dfrac{M_{\Lambda^+_c}^2-M_{12}^2-m_3^2}{2M_{12}},
\end{align}
where $m_1$, $m_2$, and $m_3$ are the masses of particles 1, 2, and 3, respectively.

On the other hand, the angular distribution of the differential decay width  in $d\Gamma/d{\rm cos}\theta$ is
\begin{equation}\label{Angle}
	d^2\Gamma = \frac{1}{(2\pi)^3}\dfrac{M_{\Lambda}}{4M_{\Lambda_c^+}}|p_{\eta}||p^*_{\pi^+}||\mathcal{T}^{\text {Total}}|^2dM_{\pi\Lambda}d{\rm cos}\theta,
\end{equation}
where $\theta$ is the angle of the outgoing $\Lambda$ (or $\pi^+$) in the center-of-mass frame of the $\pi^+\Lambda$ system, defined in Eq.~[\ref{theta}], and $|p^*_{\pi^+}|$ is the momenta of $\pi^+$ in the rest frame of the $\pi^+\Lambda$ system (Eq.~(\ref{ppi})), $|p_{\eta}|$ is the momenta of $\eta$ in the rest frame of the $\Lambda^+_c$, 
\begin{equation}\label{petastar}
	|p_{\eta}|=\dfrac{\lambda^{1/2}(M_{\Lambda^+_c}^2,m_{\eta}^2,M_{\pi^+\Lambda}^2)}{2M_{\Lambda^+_c}}.
\end{equation}
The particles' masses and widths are taken from the RPP \cite{ParticleDataGroup:2022pth}.

\section{Numerical results and discussions }\label{sec3}
\begin{figure}
    \centering
    \includegraphics[scale=0.65]{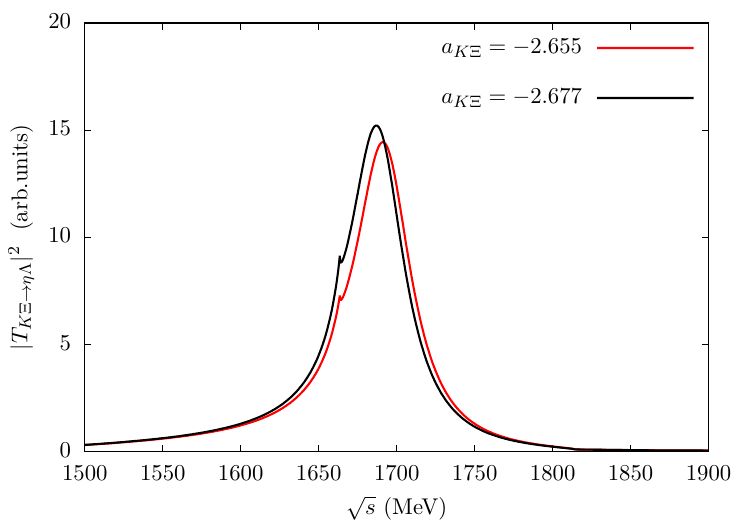}
    \caption{Modulus squared of the transition amplitudes $|T_{K\Xi\to\eta\Lambda}|^2$.}
    \label{fig:square-T}
\end{figure}

\begin{figure*}
	\subfigure[]{
		\includegraphics[scale=0.65]{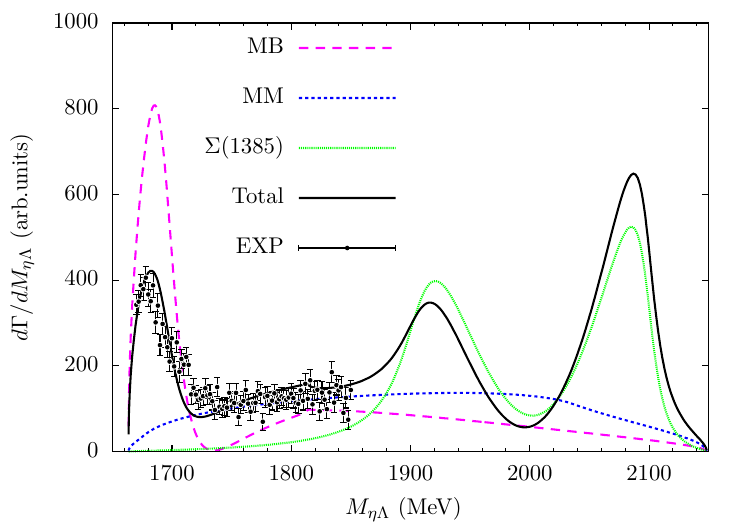}
	}
	\subfigure[]{
		\includegraphics[scale=0.65]{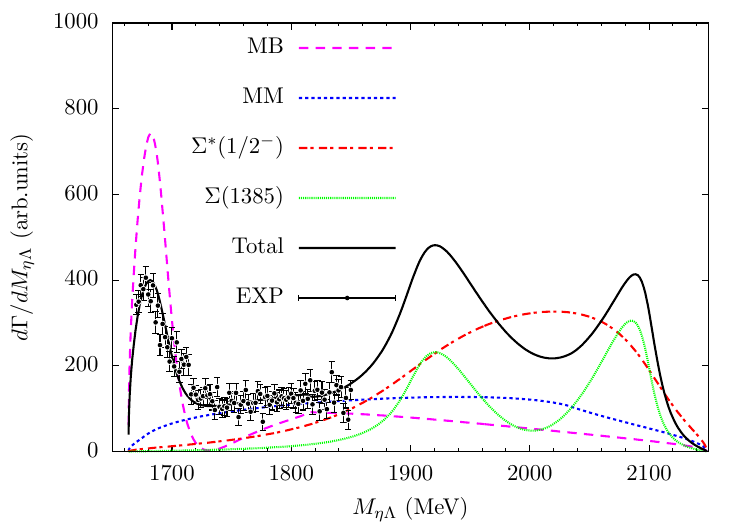}
	}
\caption{The $\eta\Lambda$ invariant mass distributions of the process $\Lambda^+_c\to \eta\pi^+\Lambda$ decay for (a) Fit~A: Without the contribution from the $\Sigma^*(1/2^-)$, and (b) Fit~B: With the contribution from the $\Sigma^*(1/2^-)$. The magenta-dashed curves show the contribution from the MB interaction, the blue-dotted curves show the contribution from the MM interaction, the red-dashed-dotted curves show the contribution from the $\Sigma^*(1/2^-)$ state, the green-dotted curves show the contribution from the $\Sigma(1385)$ state, and the black-solid curves show the results obtained with the total amplitude. The Belle data are taken from Ref.~\cite{Belle:2020xku}.}\label{fig:dwidth-etaLambda}

\end{figure*}

\begin{figure*}
	\subfigure[]{
		\includegraphics[scale=0.65]{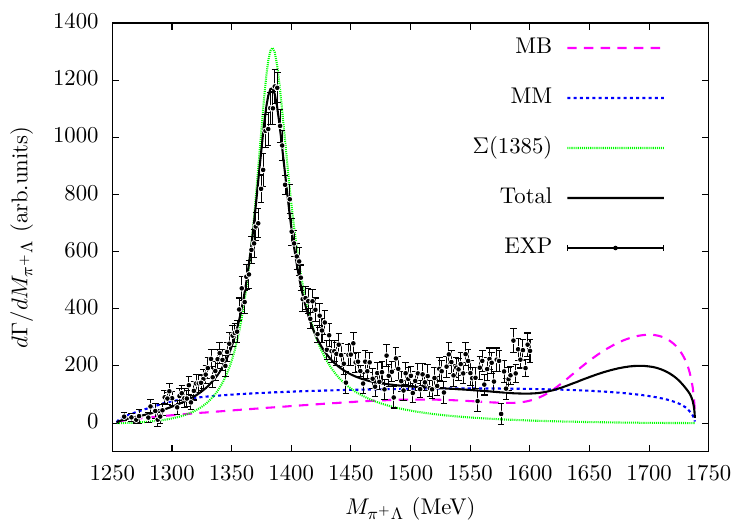}
	}
	\subfigure[]{
		\includegraphics[scale=0.65]{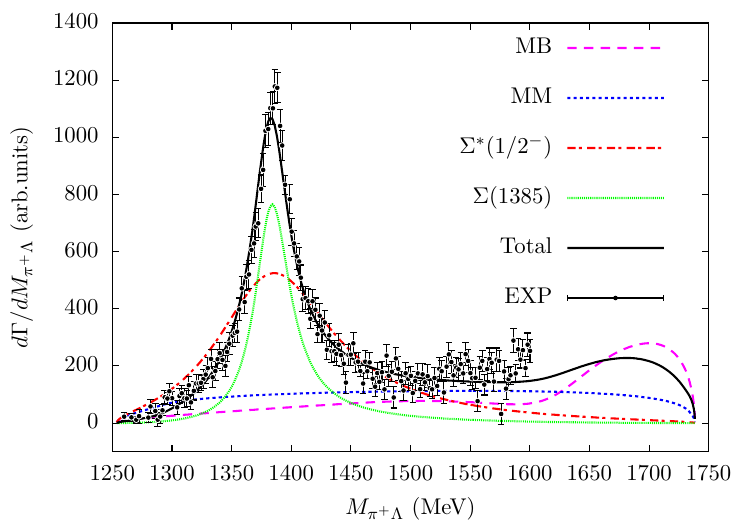}
	}
	\caption{The $\pi^+\Lambda$ invariant mass distribution of the  $\Lambda^+_c\to \eta\pi^+\Lambda$ decay for (a) Fit~A: Without the contribution from the $\Sigma^*(1/2^-)$, and (b) Fit~B: With the contribution from the $\Sigma^*(1/2^-)$. The curves have the same meaning as those of Fig.~\ref{fig:dwidth-etaLambda}.}\label{fig:dwidth-piLambda}
\end{figure*}

\begin{figure*}
	\subfigure[]{
		\includegraphics[scale=0.65]{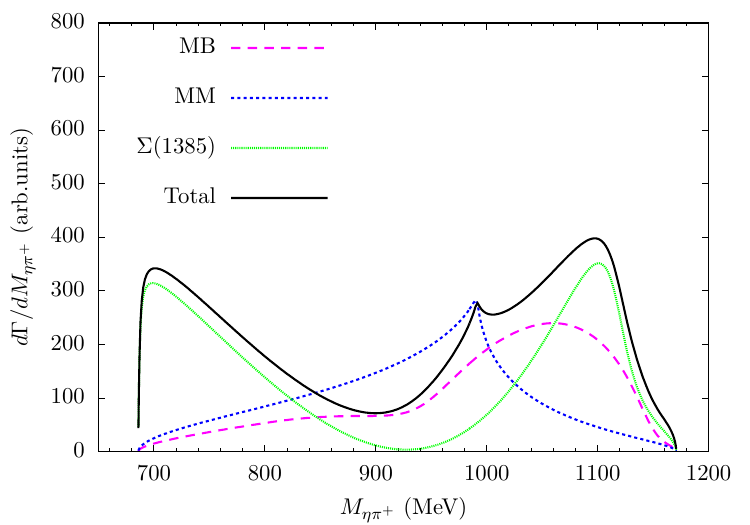}
	}
	\subfigure[]{
		\includegraphics[scale=0.65]{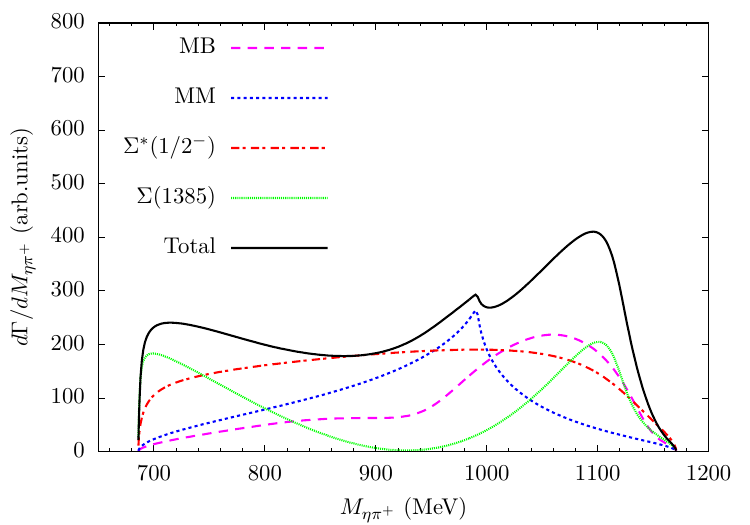}
	}
	\caption{The $\pi^+\eta$ invariant mass distribution of the $\Lambda^+_c\to \eta\pi^+\Lambda$ decay  for (a) Fit~A: Without the contribution from the $\Sigma^*(1/2^-)$, and (b) Fit~B: With the contribution from the $\Sigma^*(1/2^-)$. The curves have the same meaning as those of Fig.~\ref{fig:dwidth-etaLambda}.}\label{fig:dwidth-etapi-nocut}
\end{figure*}

\begin{figure*}
	\subfigure[]{
		\includegraphics[scale=0.65]{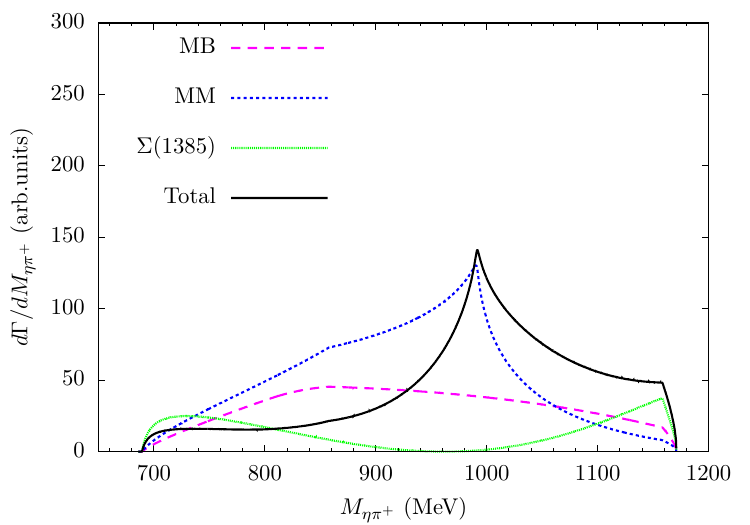}
	}
	\subfigure[]{
		\includegraphics[scale=0.65]{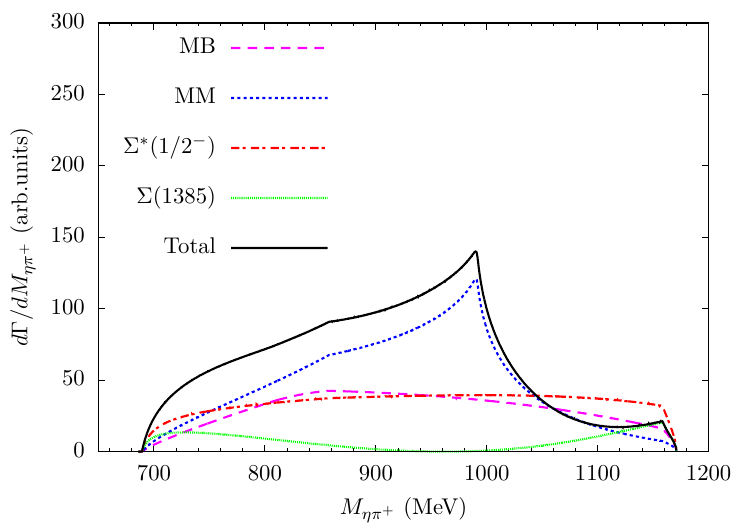}
	}
	\caption{The $\pi^+\eta$ invariant mass distribution of the $\Lambda_c\to \eta\pi^+\Lambda$ decay with the cuts of $M_{\pi\Lambda}\geq1450$~MeV and $M_{\eta\Lambda}\geq1760$~MeV  for (a) Fit~A: without the contribution from the $\Sigma^*(1/2^-)$, and (b) Fit~B: with the contribution from the $\Sigma^*(1/2^-)$. The curves have the same meaning as those of Fig.~\ref{fig:dwidth-etaLambda}.}\label{fig:dwidth-etapi-withcut}
\end{figure*}

\begin{figure*}
	\subfigure[]{
		\includegraphics[scale=0.65]{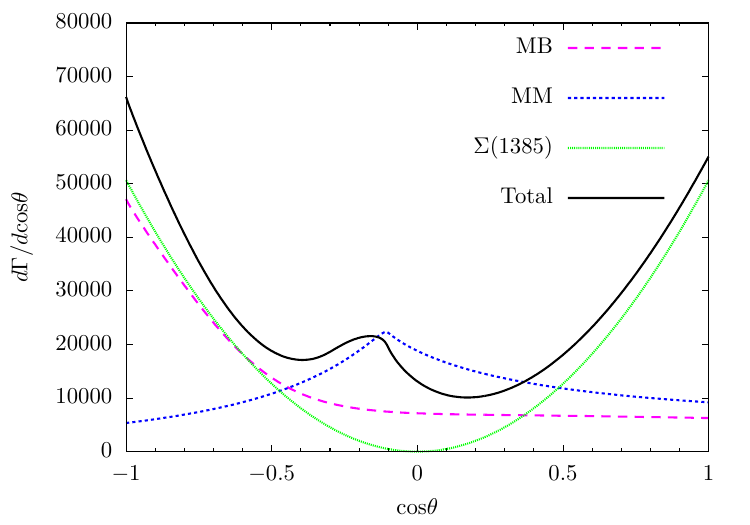}
	}
	\subfigure[]{
		\includegraphics[scale=0.65]{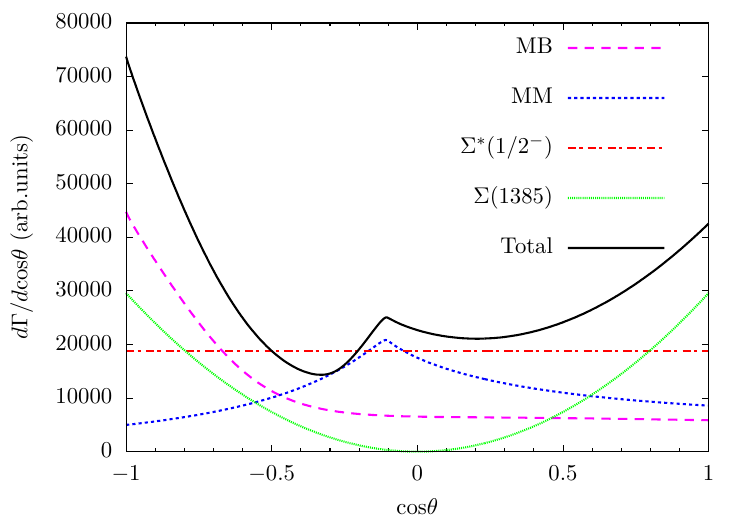}
	}
	\caption{The decay angular distributions of the  $\Lambda^+_c\to \eta\pi^+\Lambda$ decay for (a) Fit~A: Without the contribution from the $\Sigma^*(1/2^-)$, and (b) Fit~B: With the contribution from the $\Sigma^*(1/2^-)$. The curves have the same meaning as those of Fig.~\ref{fig:dwidth-etaLambda}.}\label{fig:angle-distribution-nocut}
\end{figure*}

\begin{figure*}
	\subfigure[]{
		\includegraphics[scale=0.65]{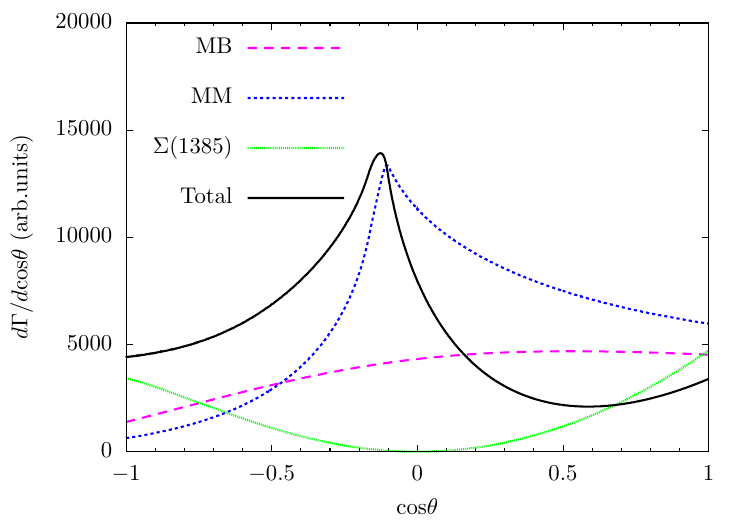}
	}
	\subfigure[]{
		\includegraphics[scale=0.65]{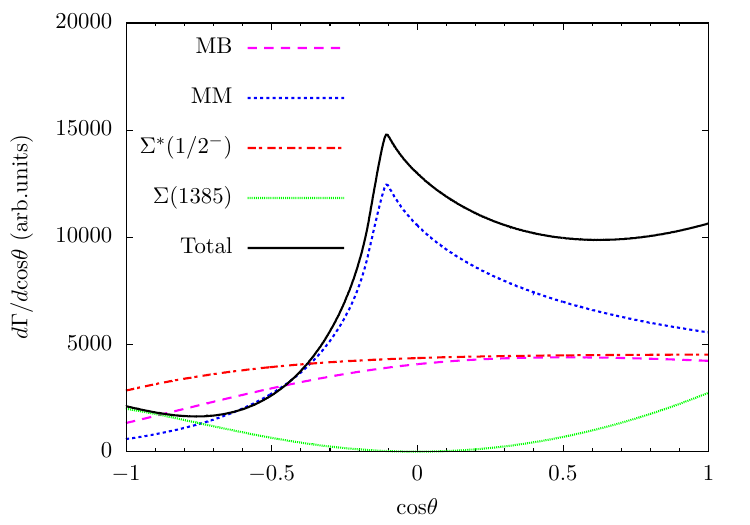}
	}
	\caption{The decay angular distributions of the $\Lambda_c\to \eta\pi^+\Lambda$ decay restricted to $M_{\pi\Lambda}\geq1450$~MeV and $M_{\eta\Lambda}\geq1760$~MeV. (a) without resonance $\Sigma(1/2^-)$, (b) with resonance $\Sigma(1/2^-)$.}\label{fig:angle-distribution-withcut}
\end{figure*}

\begin{figure*}
	\subfigure[]{
		\includegraphics[scale=0.65]{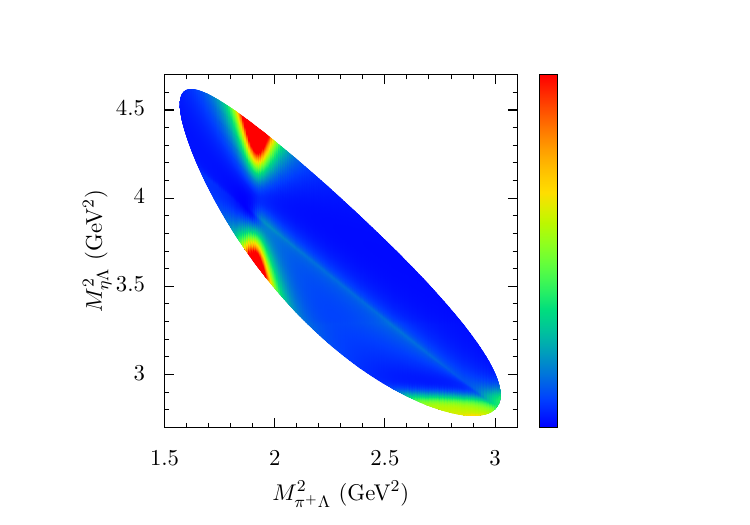}
	}
	\subfigure[]{
		\includegraphics[scale=0.65]{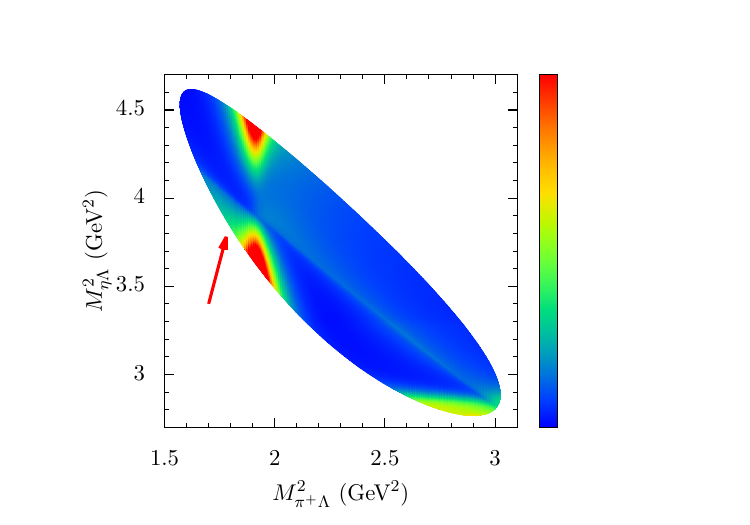}
	}
	\caption{The Dalitz plot of ``$\pi^+\Lambda$'' vs. ``$\pi^+\eta$'' in the  $\Lambda^+_c\to \eta\pi^+\Lambda$ decay for (a) Fit~A: Without the contribution from the $\Sigma^*(1/2^-)$, and (b) Fit~B: With the contribution from the $\Sigma^*(1/2^-)$.}\label{fig:dwidth-Dalitz-piLambda-etapi}
\end{figure*}

\begin{table}[htpb]
	\begin{center}	
		\caption{\label{tab:1}The fitted parameter values in this work.}
		\begin{tabular}{ccc}
			\hline\hline
		   Parameters     & Without $\Sigma^*(1/2^-)$  & With $\Sigma^*(1/2^-)$ \\
			\hline
			$V_p$      & $3.662\pm0.034$  &$3.535\pm0.031$ \\
			$V^{\Sigma(1385)}$ & $1.053\pm0.007$&$0.804\pm0.014$   \\
			$a_{K\Xi}$& $-2.655\pm0.007$ &$-2.677\pm0.007$ \\
			$\phi$&$(0.162\pm0.009)\pi $& $(0.120\pm0.013)\pi $\\
            $\phi^{\prime}$& $(1.091\pm0.021)\pi$&$(0.528\pm0.034)\pi$ \\
            $\phi^{\prime\prime}$& &$(-6.46\pm1.52)\times10^{-2}\pi$ \\
			$V^{\Sigma(1/2^-)}$& & $1.506\pm0.033$ \\
            \hline
            $\chi^2/d.o.f.$ &$3.04$  &$1.83$   \\
			\hline\hline
		\end{tabular}
	\end{center}
\end{table}

In our model, we have seven free parameters, which are, 
(1) the strength $V_p$ related to the meson-baryon interaction of Fig.~\ref{fig:MB-quark};  
(2) the subtraction constant $a_{K\Xi}$ for the $K\Xi$ channel appearing in Eq.~(\ref{G-DR});  
(3) the strength $V^{\Sigma(1385)}$ for  the contribution from $\Sigma(1385)$; 
(4) the strength  $V^{\Sigma(1/2^-)}$ for  the contribution from $\Sigma^*(1/2^-)$;
(5) three phase angles $\phi$, $\phi'$, $\phi''$ for  the interference between different contributions appearing in Eq.~(\ref{ttotal}). 

The Belle Collaboration has reported the $\eta\Lambda$ and $\pi^+\Lambda$ invariant mass distributions of the process $\Lambda_c^+\to \eta \pi^+\Lambda$~\cite{Belle:2020xku}. 
In this work, we perform two kinds of fits to the Belle measurements, (a) without the contribution from the $\Sigma(1/2^-)$ (Fit A) and (b) with the contribution from the  $\Sigma(1/2^-)$ (Fit B). 
In Fit A, we have five free parameters since the $\phi''$ and $V^{\Sigma(1/2^-)}$ are related to the $\Sigma^*(1/2^-)$. 
The fitted parameters are tabulated in Table~\ref{tab:1}.
One can see that our models of either Fit~A or Fit~B could describe the present Belle measurements. However, the $\chi^2/d.o.f.$ of Fit~B is 1.83, better than that (3.04) of Fit~A, which implies that considering the contribution from the $\Sigma^*(1/2^-)$ could improve the description of the Belle measurements.
On the other hand,  both fitted values of $a_{K\Xi}$ are close to those obtained in Refs.~\cite{Oset:2001cn,Wang:2022nac}. 
In Fig.~\ref{fig:square-T}, we have presented the modulus squared of the transition amplitude $|T_{K\Xi\to \eta \Lambda}|^2$ with $a_{K\Xi}=-2.655$ and $a_{K\Xi}=-2.677$, where one can find that both pole positions are around 1680~MeV, consistent with the RPP~\cite{ParticleDataGroup:2022pth}.

With the fitted parameters, we have calculated the invariant mass distributions of $\eta\Lambda$ and $\pi^+\Lambda$, as shown in Fig.~\ref{fig:dwidth-etaLambda} and Fig.~\ref{fig:dwidth-piLambda}, respectively.
The left and right panels correspond to the results of Fit~A and Fit~B, respectively.  
It should be stressed that two peaks appear around 1910~MeV and 2090~MeV in the $\eta\Lambda$ invariant mass distribution, which is due to the contribution from the intermediate resonance $\Sigma(1385)$, and should not be associated with any resonance.
Since the contribution from $\Sigma(1385)$ will be weakened when the $\Sigma^*(1/2^-)$ is involved, 
and the lineshape of the $\Sigma^*(1/2^-)$ is smooth,
%
it will lead to different strengths of two peak structures for Fit~A and Fit~B, which could serve as one probe to demonstrate the existence of $\Sigma^*(1/2^-)$. 

Next, we show the $\pi^+\eta$ invariant mass distributions in Figs.~\ref{fig:dwidth-etapi-nocut}(a) and \ref{fig:dwidth-etapi-nocut}(b), which correspond to the results of Fit~A and Fit~B, respectively. 
One could find a clear cusp structure around $M_{\eta\pi^+}=980$~MeV, which could be associated with the resonance $a_0(980)$. 
Indeed, the cusp structure of the scalar meson $a_0(980)$ has been observed in many processes, such as $D^+\to K_s \pi^+\eta$~\cite{BESIII:2023htx,Duan:2020vye}, $\chi_{c1}\to \eta\pi^+\pi^-$~\cite{Liang:2016hmr,BESIII:2016tqo}.
In addition, two peaks appear around the $\eta\pi^+$ threshold and $M_{\eta\pi^+}=1100$~MeV mainly due to the contribution from the $\Sigma(1385)$. 
Although the lineshape of the $\eta\pi^+$ invariant mass distributions obtained in Fit~A and Fit~B are different, it is difficult to distinguish these two cases considering the unavoidable uncertainties of future experiments.
The main reason is that the contributions from the $\Sigma(1385)$, $\Sigma^*(1/2^-)$, and $\Lambda{(1670)}$ is strong to bury the pure contribution of $a_0(980)$.

Thus, we take the cuts of $M_{\pi\Lambda}\geq1450$~MeV and $M_{\eta\Lambda}\geq1760$~MeV to eliminate the contributions from the $\Sigma(1385)$, $\Sigma^*(1/2^-)$, and $\Lambda(1670)$, and show the results in Fig.~\ref{fig:dwidth-etapi-withcut}, where the lineshapes obtained in Fit~A and Fit~B are significantly different, which could be used to test the existence of $\Sigma^*(1/2^-)$.
Such large difference is from the interference between the $a_0(980)$ and $\Sigma^*$ resonances, where in  Fit~A the $\Sigma(1385)$ is too narrow to affect the region of $M_{\pi\Lambda}\geq1450$~MeV, while in Fit~B the $\Sigma^*(1/2^-)$ still provides important contribution to $M_{\pi\Lambda}\geq1450$~MeV because of its large width.


With Eq.~(\ref{Angle}), we have calculated the angular distributions of the $\Lambda^+_c\to \eta\pi^+\Lambda$ decay with the fitted parameters of Fit~A and Fit~B, as shown in Figs.~\ref{fig:angle-distribution-nocut}(a) and \ref{fig:angle-distribution-nocut}(b), respectively. 
It is expected that there will be more events when $\pi^+$ and $\eta$ are parallel or antiparallel,  mainly due to the contributions from the $\Sigma(1385)$ and $\Lambda(1670)$. 
By performing the cuts of $M_{\pi\Lambda}\geq1450$~MeV and $M_{\eta\Lambda}\geq1760$~MeV to eliminate the contributions from the $\Sigma(1385)$, $\Sigma^*(1/2^-)$, and $\Lambda(1670)$, we present the results in Figs.~\ref{fig:angle-distribution-withcut}(a) and \ref{fig:angle-distribution-withcut}(b) for the cases of Fit~A and Fit~B, respectively. 
A clear peak appears around ${\rm cos}\theta=-0.1$ in both cases. 
However, more events are expected in the region of negative ${\rm cos}\theta$ for the case of Fit~A, while for the case of Fit~B, more events are expected in the region of positive ${\rm cos}\theta$, which could also be used to test the existence of the $\Sigma^*(1/2^-)$.
Again, it is due to the fact that, the interference between the $a_{0}(980)$ and $\Sigma^*(1/2^-)$ is large and therefore reduces the total amplitude at $\theta \sim \pi$ and $0$, respectively.

Finally, we present the Dalitz plots of `$M_{\pi^+\Lambda}$' vs `$M_{\eta\Lambda}$' with the fitted parameters of Fit~A and Fit~B in Figs.~\ref{fig:dwidth-Dalitz-piLambda-etapi}(a) and \ref{fig:dwidth-Dalitz-piLambda-etapi}(b), respectively. 
In both cases, one can find the clear signal of the $\Sigma(1385)$ around $M^2_{\pi\Lambda}=1.8\sim 2.0$ GeV$^2$, and a green band around $M^2_{\eta\Lambda}=2.8$ GeV$^2$, which should be associated with the $\Lambda(1670)$ state. 
In addition, one can easily find a thin band corresponding to the $a_0(980)$. 
It should be pointed out that more events are expected to lie in the region of  $M^2_{\Lambda\pi^+}<1.85$ GeV$^2$ and $M^2_{\eta\Lambda}<4$ GeV$^2$ when the contribution from the $\Sigma^*(1/2^-)$ is taken into account, as shown in Fig.~\ref{fig:dwidth-Dalitz-piLambda-etapi}(b).
It is in good agreement with the Belle measurements. 
Thus, one could conclude that the existence of the $\Sigma^*(1/2^-)$ is favored by the Belle measurements of the process $\Lambda^+_c\to \eta \pi^+ \Lambda$.
Unfortunately, we can not perform a fit for such Dalitz plots but only a rough comparison.
From the above discussion, we expect that the invariant mass spectrum of $\eta\pi^+$ and angular distributions, especially after the cuts, can help us establish the existence of the $\Sigma^*(1/2^-)$ particle around 1400~MeV. 

\section{ Conclusions }

Recently, the Belle Collaboration measured the process $\Lambda^+_c\to \eta\pi^+\Lambda$ and reported the $\eta\Lambda$ and $\Lambda\pi$ invariant mass distributions and the Dalitz plot, which show the signals of the resonances $\Lambda(1670)$, $a_0(980)$, and $\Sigma(1385)$. 
In this work, we have investigated this process by considering the contributions from the $S$-wave $\eta\Lambda$ and $\eta\pi$ final-state interactions, which dynamically generate the $\Lambda(1670)$ and $a_0(980)$, respectively. 
We have also considered contributions from the intermediate resonances $\Sigma(1385)$ with $J^P=3/2^+$ and the predicted state $\Sigma^*(1/2^-)$. 

We have performed two kinds of fits to the Belle measurements of the $\eta\Lambda$ and $\pi\Lambda$ invariant mass distributions,
without (Fit~A) and with (Fit~B) including the contribution of the $\Sigma^*(1/2^-)$.
%
Although our results obtained in both cases could describe the Belle measurements of the $\eta\Lambda$ and $\pi\Lambda$ invariant mass distributions, the $\chi^2/d.o.f.$ of Fit~B is 1.83, better than that (3.04) of Fit~A, which implies that considering the contribution from the $\Sigma^*(1/2^-)$ could improve the descriptions of the Belle measurements.
But, to be fair, the peaks in the invariant mass of the $\eta\Lambda$ and $\pi\Lambda$ can be well-explained by both fits. 

Then, we predicted the $\eta\pi^+$ invariant mass distributions and found one clear peak structure around $980$~MeV, which should be associated with the scalar meson $a_0(980)$.
By performing the cuts of $M_{\pi^+\Lambda}\geq1450$~MeV and $M_{\eta\Lambda}\geq1760$~MeV to eliminate the contributions from the  $\Sigma(1385)$, $\Sigma^*(1/2^-)$, and $\Lambda(1670)$, the line shapes of the $\eta\pi^+$ invariant mass distributions are significantly different when considering the contribution from the $\Sigma^*(1/2^-)$. 
Furthermore, we have also predicted the angular distribution for this process and found that there will be more events when $\pi^+$ and $\eta$ are parallel or antiparallel, which is mainly due to the contributions from the $\Sigma(1385)$ and $\Lambda(1670)$.
By performing the cuts of $M_{\pi^+\Lambda}\geq1450$~MeV and $M_{\eta\Lambda}\geq1760$~MeV to eliminate the contributions from the  $\Sigma(1385)$, $\Sigma^*(1/2^-)$, and $\Lambda(1670)$, one finds more events in the region of positive ${\rm cos}\theta$.
All of these differences stem from the interference between the  $a_0(980)$ and $\Sigma^*(1/2^-)$; thus, it provides a unique place to identify the existence of $\Sigma^*(1/2^-)$ around 1.4~GeV.

With the fitted parameters in both cases, we have calculated the Dalitz plots and found clear signals of the $\Lambda(1670)$, $\Sigma(1385)$, and $a_0(980)$.
It should be pointed out that, with the contribution from the $\Sigma^*(1/2^-)$, more events are expected to lie in the region of  $M^2_{\Lambda\pi^+}<1.85$~GeV$^2$ and   $M^2_{\eta\Lambda}<4$~GeV$^2$, which are in agreement with  the Belle measurements. 
Thus, one could conclude that the existence of the $\Sigma^*(1/2^-)$ is favored by the Belle measurements of the process $\Lambda^+_c\to \eta \pi^+ \Lambda$, and our predictions could be helpful for future experimental analysis of this process.

\section*{Acknowledgments}
We would like to acknowledge the fruitful discussions with Professor Eulogio Oset. 
L.S.G and J.J.X are partly supported by the National Key R\&D Program of China under Grant No.~2023YFA1606700. E.W is partly supported by the National Key R\&D Program of China under Grant No.~2024YFE0105200.
This work is supported by the Natural Science Foundation of Henan under Grant No.~232300421140 and No.~222300420554, the National Natural Science Foundation of China under Grants No.~12475086, No.~12192263, No.~12205075, No. 12175239, No.~12221005, No.~12075288, No.~12361141819,
 the Open Project of Guangxi Key Laboratory of Nuclear Physics and Nuclear Technology, No.~NLK2021-08, 
and the Central Government Guidance Funds for Local Scientific and Technological Development, China (No.~Guike ZY22096024),
and also by the National Key Research and Development Program of China under Contract No.~2020YFA0406400, 
and also by the Chinese Academy of Sciences under Grant No.~YSBR-101, 
and also by the Youth Innovation Promotion Association CAS.

\end{document}